\documentclass[twocolumn,superscriptaddress,floatfix]{revtex4-2}

\usepackage{graphicx}
\usepackage[para]{threeparttable}
\usepackage{amsmath}
\usepackage{lineno}
\usepackage{booktabs}
\usepackage{bbold}
\usepackage{bbm}
\usepackage{tabularx}
\begin{document}

\title[]{Robust extrapolation in nuclear mass predictions using physics-related activation functions}

%Robust extrapolation using physics-related activation functions in neural networks for nuclear masses
%Physics-related activation functions enable robust extrapolation in neural networks for nuclear masses
%Robust extrapolation in neural networks for nuclear mass predictions

\author{C.~H.~\surname{Kim}}
\affiliation{Department of Physics, Sungkyunkwan University, Suwon 16419, Republic of Korea}

\author{K.~Y.~\surname{Chae}}
\email{kchae@skku.edu}
\thanks{Fax: +82-31-290-7055}
\affiliation{Department of Physics, Sungkyunkwan University, Suwon 16419, Republic of Korea}

\author{M.~S.~\surname{Smith}}
\thanks{Present Address: Stellar Science Solutions}
\affiliation{Physics Division, Oak Ridge National Laboratory, Oak Ridge, Tennessee 37831, USA}

\begin{abstract}

Given the importance of nuclear mass predictions, numerous models have been developed to extrapolate the measured data into unknown regions. While neural networks---the core of modern artificial intelligence---have been recently suggested as powerful methods, showcasing high predictive power in the measured region, their ability to extrapolate remains questionable. This limitation stems from their `black box' nature and large number of parameters entangled with nonlinear functions designed in the context of computer science. In this study, we demonstrate that replacing such nonlinear functions with physics-related functions significantly improves extrapolation performance and provides enhanced understanding of the model mechanism. Using only the information about neutron ($\textnormal{\textit{N}}$) and proton ($\textnormal{\textit{Z}}$) numbers without any existing global mass models or knowledge of magic numbers, we developed a highly accurate model that covers light nuclei ($\textnormal{\textit{N}}$, $\textnormal{\textit{Z}}$ $>$ 0) up to the drip lines. The extrapolation performance was rigorously evaluated using the outermost nuclei in the measurement landscape, and only the data in the inner region was used for training. We present details of the method and model, along with opportunities for future improvements.

\end{abstract}

\maketitle

\section{Introduction}

Nuclear masses play a critical role in various fields, including not only nuclear physics, but also astrophysics, neutrino studies, nuclear energy, nuclear security, and more \cite{D.Lunney2003}. Numerous studies have been extensively done to predict the unknown masses from existing measurements \cite{W.Myers1974, J.Duflo1995, H.Koura2005, N.Wang2014, P.Moller2016}. Recently, the use of neural networks has emerged as a viable method for accurate predictions across fitting data space \cite{R.Utama2016, Z.Niu2018, A.Lovell2022, M.Mumpower2022, A.Boehnlein2022}. Still, the ability to extrapolate beyond the fitting data space remains limited due to the inherent characteristics of neural networks \cite{E.Barnard1992, P.Haley1992, G.Martius2016, Y.Gal2018, C.Kim2024}. While the power of (deep) neural networks originates from the large number of parameters entangled with activation functions, this feature can undermine effectiveness outside the training data distribution as they easily become over-parametrized and overfitted to the specific data space \cite{C.Louizos2018, M.Bejani2021}. This drawback also gives rise to the `black box' nature of deep neural networks \cite{V.Buhrmester2021, L.Jospin2022}. 

Many efforts have been made to establish methods that find the true mathematical expression underlying the given data, which will naturally extrapolate well beyond the data \cite{M.Schmidt2009, G.Martius2016, S.Sahoo2018, S.Udrescu2020}. A method called equation learning was suggested for this purpose: it replaces activation functions with scientific functions (e.g., logarithm) that could be components of the underlying expression and uses sparse learning that prunes unimportant parameters \cite{G.Martius2016, S.Sahoo2018, M.Werner2021}. 

Typical activation functions in computer science tasks are rectified linear unit (ReLU), hyperbolic tangent, and others. They have been empirically proven to give stable and fast training of deep neural networks and high performance on various tasks such as computer vision, natural language processing, and many more \cite{A.Geron2019, K.Murphy2022}. However, because they were chosen for such a purpose in computer science, they are not easily interpretable for physics, and this is exacerbated by the large number of model parameters. These characteristics also contribute to the weakness of deep neural networks in extrapolating predictions beyond their training range. The replacement of the conventional activation functions with functions regularly used in the specific physics task could possibly mitigate this issue and provide better prediction accuracy. Furthermore, the number of parameters should be reduced to alleviate over-parametrization and to achieve a less complex structure. The reduction can be carried out using a method known as sparse learning with the $L_0$ regularization \cite{C.Louizos2018}.

In ideal equation learning, the analytical expression of the neural network can be obtained after extensively pruning the network with replaced activation functions. This idea was demonstrated on ``toy model'' examples, but few successful applications were performed for complex and high-dimensional physics tasks \cite{G.Martius2016, S.Sahoo2018, S.Kim2021}. In practice, most physics systems might not be represented by a simple equation. Additionally, applying strong regularization to obtain a simple equation does not necessarily yield optimal performance.

\begin{figure}[t]
\includegraphics[width=0.48\textwidth]{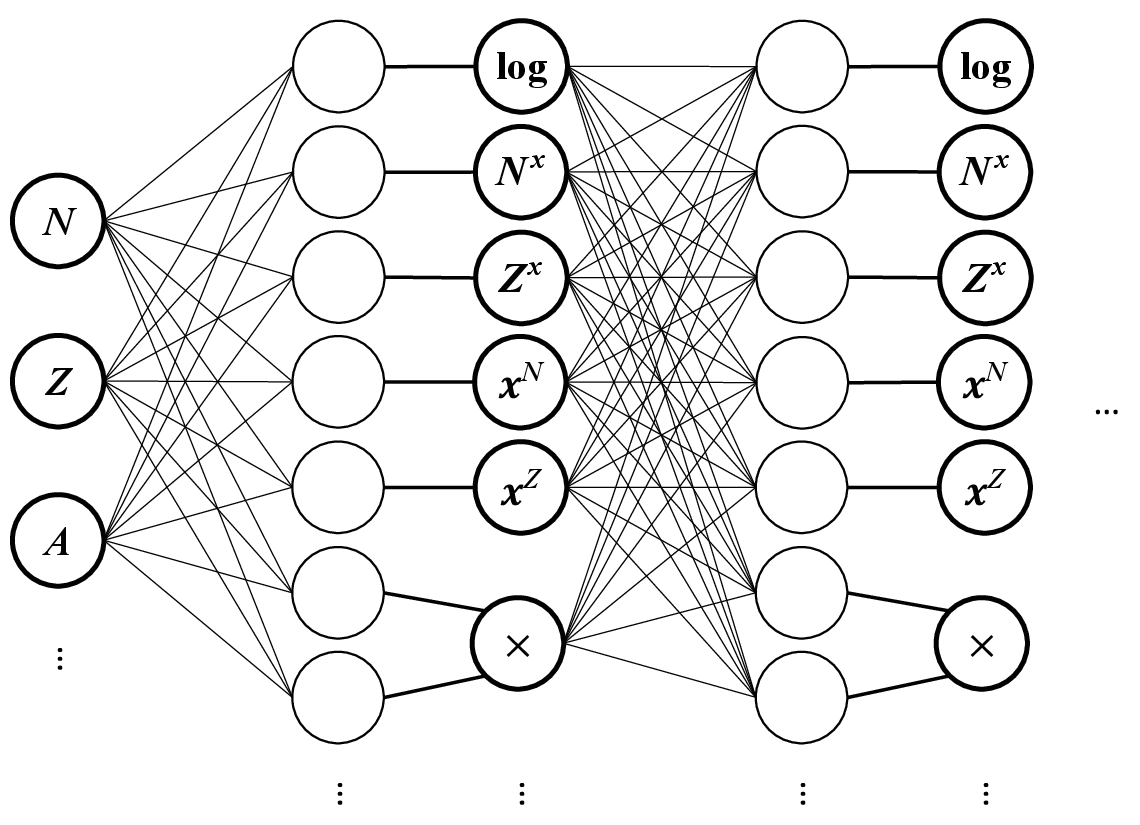}
\caption{A sketch of the PAF network. Each node of $\div$ or $\times$ takes two nodes as it is operation for two values. $f_{\text{N,Z}}$ represents functions composed of $N$ and $Z$.}
\label{fig_EQL}
\end{figure}

In this study, we highlight that using the physics-related activation functions (PAF) with a modest level of sparse learning can lead to an accurate model with highly improved extrapolation---without pursuing a simple analytical equation. We developed a mass model capable of effectively extrapolating beyond existing data, covering nuclei from light isotopes ($N,Z>0$) to the drip lines, with information only on neutron and proton numbers. Fig.~\ref{fig_EQL} shows a sketch of the PAF network structure customized for this study. Its reliability was thoroughly evaluated by allocating a large and less biased portion of experimental data for validation and test. We found that such a network produces highly accurate predictions for both interpolation and extrapolation compared to conventional neural networks and existing global mass models. Throughout the paper, we present details of the network and training method, the performance compared to the conventional neural networks, and a discussion on potential improvements to inspire future advances.

\section{Neural network of physics-related activation functions}

Table~\ref{tab_input_function} summarizes the inputs of the model. Only the information on $N$ and $Z$ was used for inputs; we specifically avoided using any predictions of existing global mass models or magic number information in order to find out if the PAF network could naturally find such characteristics. In addition to the plain integers $N$ and $Z$, odd-even information and various combinations of $N$ and $Z$ were also included as inputs, which may help characterize nuclear mass systematics.

\begin{table}[t]
\caption{Inputs and activation functions of the PAF network. $x$ represent the values propagated through the nodes, and $\%$2 represents the remainder of a division by 2.}\label{tab_input_function}
\newcommand{\colwidth}{1.28cm}
\begin{tabular}{p{\colwidth}p{\colwidth}p{\colwidth}|p{\colwidth}p{\colwidth}p{\colwidth}}
%\begin{tabularx}{0.48\textwidth}{XXX|XXX}
\multicolumn{3}{c|}{Inputs} &\multicolumn{3}{c}{Activation functions} \\
\hline
$N$ & $Z$ & $A$ & $\mathbb{1}$ & $\times$ & $\div$ \\
$1/N$ & $1/Z$ & $1/A$ & log($x$) & sin($x$) & ReLU \\
$N\%$2 & $Z\%$2 & & $N^x$ & $Z^x$ & \\
$N/Z$ & $Z/N$ & & $x^N$ & $x^Z$ & \\
$A^{1/3}$ & $A^{2/3}$ & & \multicolumn{2}{l}{$((N-Z)/A)^x$} & \\
\hline
\end{tabular}
\end{table}

%Increasing decreasing?

As we are not aware of the specific functions that can capture the underlying physics of the mass system, we tested numerous functions and their combinations. Table~\ref{tab_input_function} shows a list of the functions found to be effective. The feed-forward (densely connected) neural networks already contain the addition ($+$) and subtraction ($-$) in their matrix multiplications with weights and biases. The multiplications ($\times$) and divisions ($\div$) were included to complete the basic arithmetic operations. Commonly used functions in physics, such as sinusoidal and logarithmic functions, were also used. Additionally, we implemented diverse functions composed of $N$ and $Z$, which clearly improved the extrapolation performance. The outputs that may not require any additional operation can pass the identity operation ($\mathbb{1}$). 
%More details are described in Methods.

Sparsity in a neural network can be induced using the $L_0$ norm regularization \cite{G.Martius2016, C.Louizos2018}. The $L_0$ regularization in the loss function $\mathcal{L}$ penalizes parameters $\boldsymbol{\theta}$ for being non-zero values: 
%\begin{linenomath}
\begin{align}
\begin{split}
\mathcal{L} = -\text{log}(p(\boldsymbol{y} \!\mid\! \boldsymbol{x}, \boldsymbol{\theta}))+\lambda \|\boldsymbol{\theta}\|_0,
\label{eq_L0}
\end{split} \\
\begin{split}
\|\boldsymbol{\theta}\|_0 = \sum_i \mathbbm{I}[\theta_i \neq 0],
\label{eq_L0_2}
\end{split}
\end{align}
%\end{linenomath}
where $-\text{log}(p(\boldsymbol{y} \!\mid\! \boldsymbol{x}, \boldsymbol{\theta}))$ is the negative log likelihood with input-output pairs $\boldsymbol{x}$ and $\boldsymbol{y}$, and $\|\boldsymbol{\theta}\|_0$ represents the number of non-zero parameters with a weighting factor $\lambda$. This regularization is different from $L_1$ or $L_2$ as it does not penalize the magnitude of values. The non-differentiable issue of this $L_0$ norm was solved by using stochastic gates based on the hard concrete distribution; see Ref. \cite{C.Louizos2018} for details.

Every 20 epochs during the training, an additional two steps were adopted besides the optimization of the loss function. First, the bound penalty suggested by Ref. \cite{S.Sahoo2018} was used to bound the magnitude of predictions on the extrapolation area in a reasonable range to avoid its drastic changes during the training: 
\begin{equation}
\mathcal{L}_{\text{bound}} = \sum_i \text{max}(\mid\!f_{\text{NN}}(\boldsymbol{x_i}; \boldsymbol{\theta})-B_1\!\mid\!-B_2,0),
\label{eq_Lbound}
\end{equation}
where the samples $\boldsymbol{x_i}$ were randomly selected from the possible extrapolation region ($1, 1\leq N, Z \leq 300, 200$). Specifically, the bound range of mass excesses was set to [$-$0.1, 2] $u$ (atomic mass unit) by defining $B_1$, $B_2$=0.95, 1.05 $u$. While this range was tuned to find the optimal setting, the experimentally observed lowest mass excess, $\sim$0.098 $u$ (from $^{118}$Sn), was considered to set the lower bound. 
% max and bar definition

The second step steers the model prediction to softly meet the (algebraic) Garvey-Kelson (GK) relation \cite{G.Garvey1966}. Similar to the bound penalty, the loss for the second step was defined as:
\begin{align}
\begin{split}
\mathcal{L}_{\text{GK}} = \sum_i \text{max}(\mid\!\Delta(\boldsymbol{x_i}; N,Z)\!\mid\!-G,0), 
\label{eq_LGK_1}
\end{split} \\
\begin{split}
&\Delta(\boldsymbol{x}; N,Z) \equiv M_{\text{pred}}(N,Z)+M_{\text{pred}}(N+1,Z-2) \\
&+M_{\text{pred}}(N+2,Z-1) - M_{\text{pred}}(N+2,Z-2) \\
&-M_{\text{pred}}(N,Z-1)-M_{\text{pred}}(N+1,Z),
\label{eq_LGK_2}
\end{split}
\end{align}
where $\boldsymbol{x_i}$ were also randomly selected from the extrapolation region, and $M_{\text{pred}}(N,Z)$ is the model prediction. While $\Delta(\boldsymbol{x}; N,Z)$ is equal to zero according to the GK relation, it is not an exact formula \cite{D.Lunney2003}. The use of Eq.~\ref{eq_LGK_1} can avoid forcing the model predictions to strictly follow the relation. The range of deviations from the relation was set to [$-$0.5, 0.5] $u$ by defining $G$=0.5 $u$. 
%See Methods for more details of these two penalties. 

\section{Data preparation}

Achieving effective extrapolation requires proper validation to ensure model reliability. For machine learning based methods, proper validation necessitates non-biased statistical tests over the landscape both for inter- and extrapolation, rather than using all (or most) existing data for fitting and a small subset for testing. This rigorous validation is critical for nuclear property predictions, as individual isotopes possess distinctive features, making the evaluation dataset prone to bias. 

The experimental data in the current study were taken from the atomic mass evaluation ($\texttt{AME2020}$) \cite{AME2020}. The data was divided into five subsets: the training dataset, two validation datasets for inter- and extrapolation, and two test datasets for inter- and extrapolation. Both validation and test datasets should be prepared to avoid selecting a model biased on the validation datasets during iterative fine-tuning of the model. The extrapolation datasets comprised the outermost nuclei in the nuclear landscape. These include the nuclei with the lowest and highest $Z$ in each isotonic line and those with $N$ in each isotopic line. The training data were sampled exclusively from the interpolation region. Out of the total 2,548 available mass data points, 1,736 were allocated for training, 435 for evaluating interpolation performance, and 377 for evaluating extrapolation performance. 

%Approximately 68$\%$ of the data were used for training, 17$\%$ for evaluating interpolation performance, and 15$\%$ for extrapolation.

\section{Training details}

We used the mean squared error for the loss function with the Adam optimizer \cite{Adam}. The sparsity was kept moderate to obtain high accuracy and extrapolation strength. During the tuning of the network details, such as choices of activation functions, we monitored the root mean square (RMS) errors on the validation datasets as well as the formation of drip lines. 

\begin{figure}[t]
\includegraphics[width=0.48\textwidth]{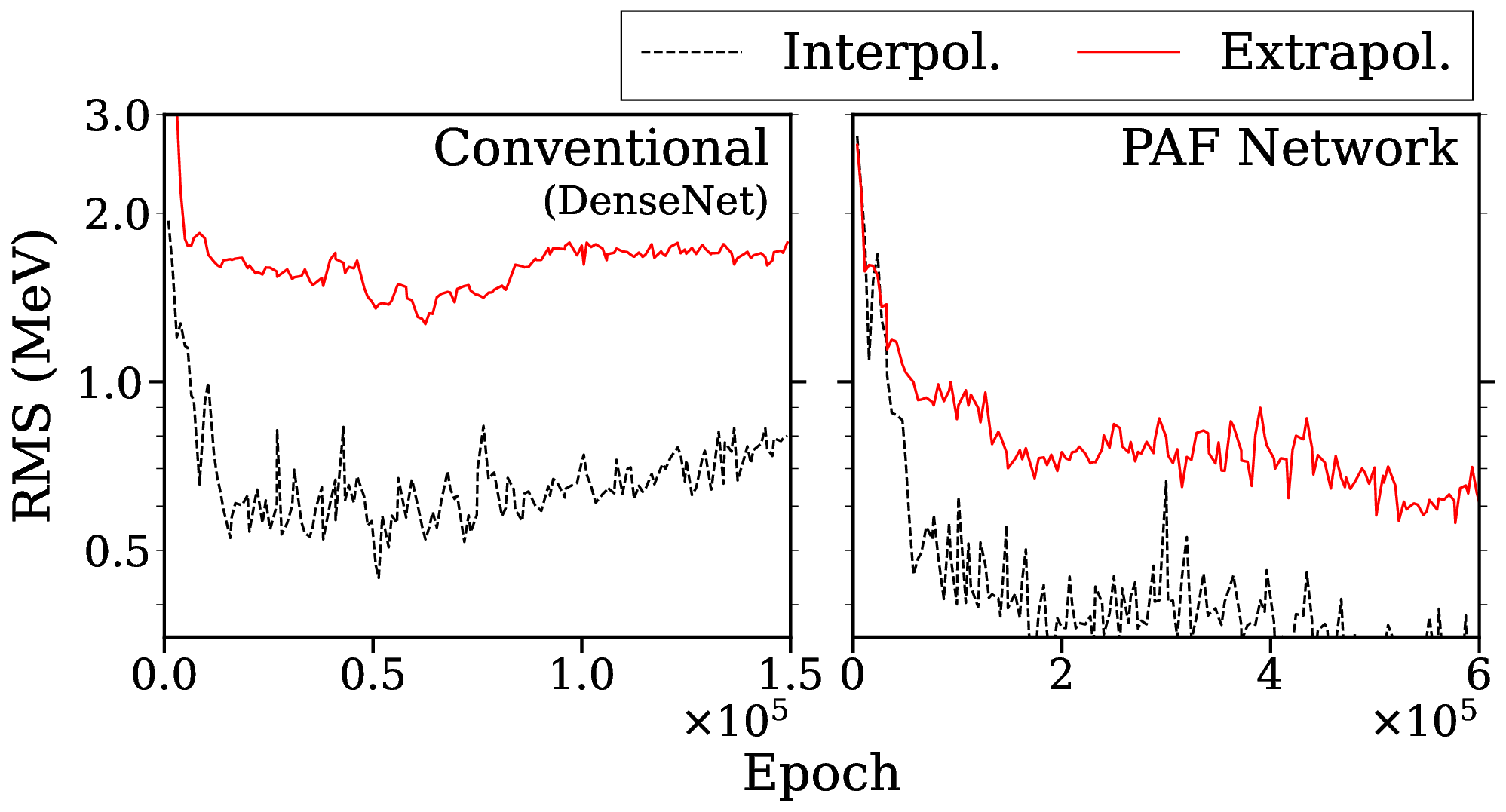}
\caption{Learning curves of the conventional neural network and PAF network. The curves represent the RMS errors on the validation datasets for interpolation and extrapolation during training. To smooth the curves, an epoch with the lowest RMS for extrapolation is shown for every few thousand epochs.}
\label{fig_curves}
\end{figure}

\begin{table}[t]
\caption{Root mean square errors of different models. It presents the evaluation results on the combined validation and test datasets. Since the Duflo-Zuker, KTUY, WS4, and FRDM were fitted using the different $\texttt{AME}$ version \cite{J.Duflo1995, H.Koura2005, N.Wang2014, P.Moller2016}, a clear comparison of its inter- and extrapolation performance with the current model is not feasible; therefore, we only present its total RMS error on the $\texttt{AME2020}$ data.}\label{tab_RMS}
\newcommand{\colwidth}{2.40cm}
\newcommand{\colwidths}{1.84cm}
\begin{tabular}{p{\colwidth}p{\colwidths}p{\colwidths}p{\colwidths}}
%\begin{tabular}{llll}
% Include training dataset in the interpol? 
% Include FRDM, etc. too? 
& \multicolumn{3}{c}{RMS ($N,Z\geq8$) / ($N,Z>0$) (keV)} \\ \cmidrule{2-4}
Model & Interpol. & Extrapol. & Total \\
\toprule
PAF Network & 207 / 255 & 308 / 396 & 258 / 328 \\
%w/o $[N^x$, $Z^x]$ & 209 / 257 & 337 / 451 & 274 / 360 \\
%w/ $[x^{(N-Z)/A}]$ & 240 / 283 & 366 / 463 & 304 / 377 \\
\toprule
DenseNet & 494 / 532 & 1048 / 1173 & 795 / 889 \\
ConvNet & 318 / 409 & 531 / 658 & 428 / 539 \\
\toprule
Duflo-Zuker \cite{J.Duflo1995} & - & - & 428 / - \\
KTUY \cite{H.Koura2005} & - & - & 733 / 743\footnotemark[1] \\
WS4 \cite{N.Wang2014} & - & - & 295 / - \\
FRDM \cite{P.Moller2016} & - & - & 606 / - \\
\toprule
\end{tabular}
%\item [1] $N,Z\geq2$.
\footnotetext[1]{$N,Z\geq2$.}
\end{table}

The idea of deep ensembles was implemented to enhance the predictive performance and estimate model uncertainty (see Refs.~\cite{B.Lakshminarayanan2017, C.Kim2024}). Sixteen models were trained using the same network and training method with random initialization of model parameters. Aggregating predictions of these models gave a better predictive performance. Still, because of penalty epochs and non-typical regularization, the methodology of deep ensembles might not be strictly followed, possibly resulting in uncertainties that are less well-calibrated. Additional studies on using deep ensembles for such cases are required in the future. Here, we used the standard deviation of the prediction ensembles as an approximate measure of model uncertainties.

Various conventional deep neural networks (DNNs) were also trained to compare with the PAF model. These include densely connected neural networks (DenseNet) with different activation functions and regularizations, as they have the same network structure with the current PAF model. Additionally, we trained convolutional neural networks (ConvNet) to examine the dependence of the results on the network structure. The DenseNet and ConvNet models using ReLU activation achieved the lowest RMS errors and were therefore chosen as representatives. 
%See Methods for more details. 

%learning curve
%RMS
%Landscape - Sep
%Drip lines
%(model interpretation)

\begin{figure*}[!t]
%\centering
%\includegraphics[width=0.65\textwidth]{Landscape_comp.eps}
\includegraphics[width=1\textwidth]{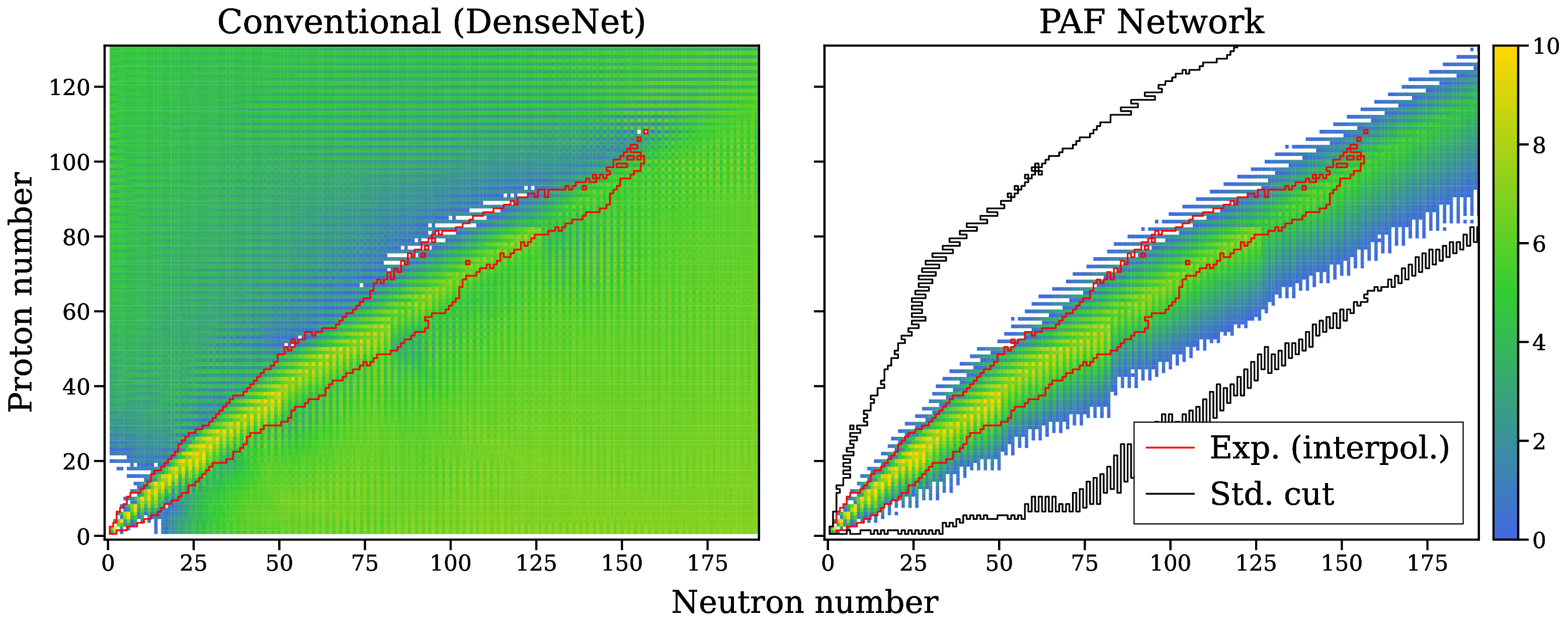}
% DNN vs PAF 
% Sep E / unc? / training data distribution / FRDM / known drip line
% maybe with embedded plot (zoom in the half-lives part)
\caption{Nuclear landscapes for separation energies. For each nucleus, the lower energy between neutron and proton separation energies is shown, excluding the nuclei with separation energies less than zero, i.e., unbound. The red line shows the experimental data area allocated for interpolation. The outer regions of the landscape for the PAF network were trimmed based on a cutoff using standard deviations (std.) from the model ensembles.}
\label{fig_landscape}
\end{figure*}

\section{Results}

Fig.~\ref{fig_curves} shows the learning curves of the conventional DNN and PAF models. The error of the PAF model on the validation dataset for extrapolation was clearly reduced over epochs, whereas the DNN could not effectively predict the extrapolation data. The RMS errors of the datasets for each model are shown in Table~\ref{tab_RMS}. Among all training epochs, we selected the DNN and PAF models that achieved the lowest RMS errors on the validation dataset for extrapolation. We note that interpolation errors could be further reduced---at the expense of extrapolation performance---by selecting models based on the minimum RMS errors on the interpolation dataset, as is conventionally done. The significantly lower RMS errors of the PAF model indicate that the current method clearly improves the extrapolation ability and could be well-suited for deep learning applications in physics research. 

%with and without certain activation functions, sparse learning, bp, GK

\begin{figure}[!t]
\includegraphics[width=0.48\textwidth]{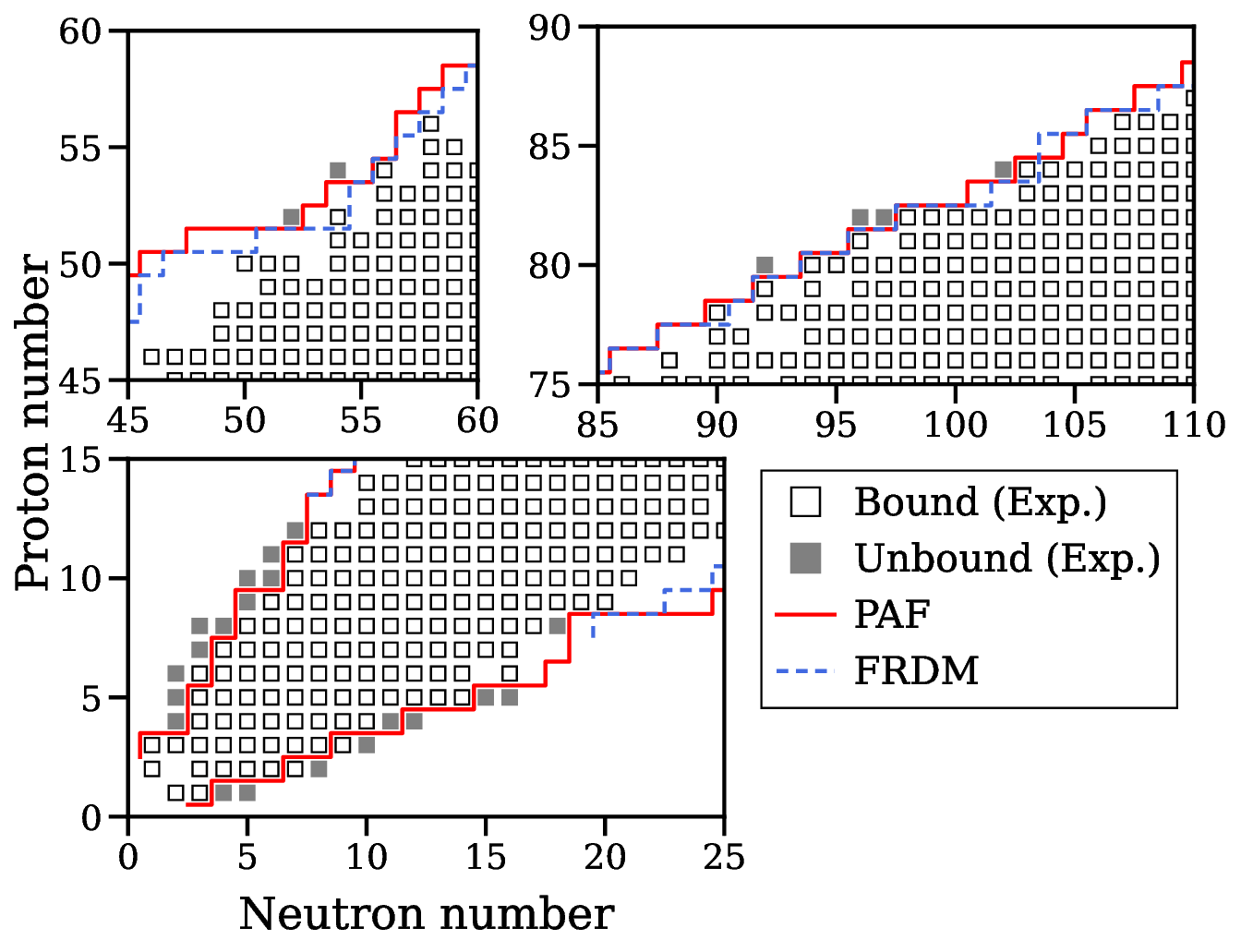}
\caption{Experimental mass data and two nucleon drip lines. Each nucleus is labeled as either bound or unbound based on the experimental data.}
\label{fig_driplines}
\end{figure}

Fig.~\ref{fig_landscape} shows the separation energy predictions of the DNN and PAF models across the nuclear landscape. While both models perform well within the interpolation region, the DNN does not exhibit physically consistent behavior beyond the data range, and does not predict the existence of drip lines. DNNs with different activation functions, structure, and regularization also gave no significant improvements in the extrapolation regions. In contrast, the PAF model successfully forms both proton and neutron drip lines. Even though most nuclei near the drip lines were excluded from the training dataset, as shown in Fig.~\ref{fig_driplines}, the drip lines formed by the PAF model agree well with the experimental data. 

%In the PAF model predictions, there are two notable features on the neutron-rich side. First, the effect of shell quenching is predicted at $N=126$, which can be seen in Extended Data Fig.~\ref{fig_shellgaps}. While there is a certain discrepancy between the experimental data and predictions from the PAF for the neutron shell gap near the $N=126$ and $Z=82$, the trend of the data appears to be reasonably reproduced by the predictions. Second, the neutron drip line is closer to the line of stability than other mass models, such as FRDM. This feature emerges for $N>82$, as shown in Extended Data Fig.~\ref{fig_rpath}, which will affect the rapid neutron capture process path which approximately follows the contours of neutron separation energies. 

% Still, the PAF model predictions for even higher mass regions were filtered out because of the increasing standard deviations of the predictions. 

\begin{figure}[!t]
\includegraphics[width=0.48\textwidth]{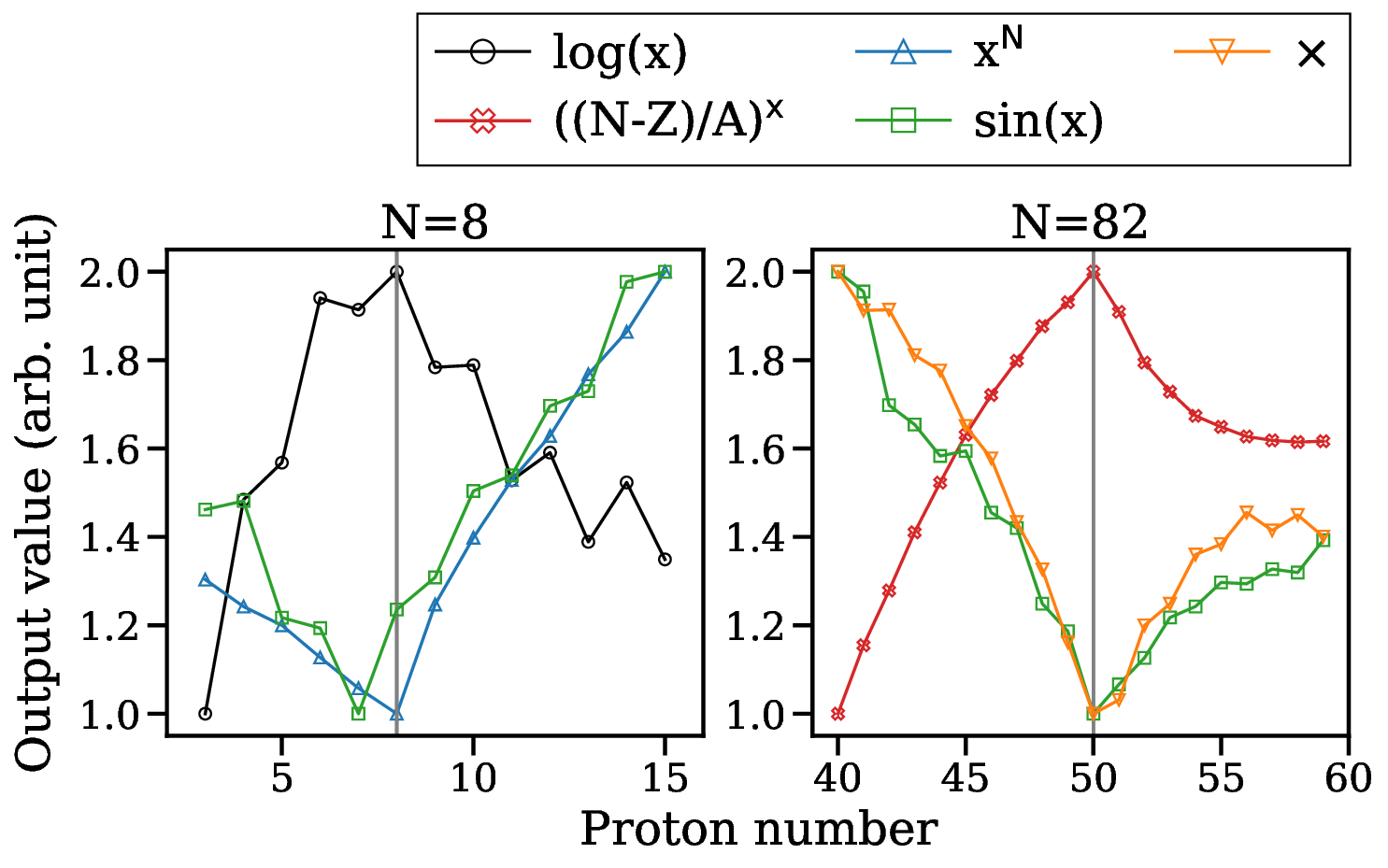}
\caption{Outputs of the activation functions before the last layer for various $N$ and $Z$. As the output ranges of the activation functions differ, all outputs were normalized to an arbitrary unit for consistent comparison. Gray bars represent magic numbers. Only activation functions exhibiting readily notable features are presented for each subplot. %See Methods for more details.
}
\label{fig_interpret_N}
\end{figure}

The PAF network with an additional function $x^{(N-Z)/A}$ or without $N^x$ and $Z^x$ leads to increased RMS errors, approximately 10 to 20$\%$ for the extrapolation. These results indicate that the nuclear mass system may be more closely associated with terms like $N^x$ and $Z^x$, or comparable functions of these, rather than $x^{(N-Z)/A}$. The role of each activation function could be interpreted by analyzing its output at each layer. Fig.~\ref{fig_interpret_N} shows the outputs of each activation function before the last layer as $Z$ varies while $N$ is fixed at magic numbers. Certain functions clearly show inflection points and other discontinuous behaviors at magic numbers, which suggest that these (or comparable) functions might be related to the emergence of magic number effects in nuclear masses. It should be noted that these functions do not solely produce such behaviors, as the outputs from previous layers are accumulated and propagated as the inputs to them. 

%The pruning of weights may reflect the relative importance of certain activation functions, if the weights connected to a certain activation function are pruned more or less frequently. While we found that the weights connected to ReLU were notably more zeroed, this may simply indicate that ReLU is more susceptible to the regularization due to its behavior of outputting zero for negative inputs. 

%interpretation analysis

For evaluation of extrapolation performance, it may seem preferable to use newly updated experimental data from a recent atomic mass evaluation as the validation and test datasets, while using the remaining data for interpolation. However, this approach results in a biased and limited number of samples. We trained the same DNN and PAF networks with the extrapolation datasets composed of newly updated data in $\texttt{AME2020}$ relative to $\texttt{AME2012}$. RMS errors ($N,Z>0$) of inter- and extrapolation for ConvNet are 426 and 446 keV, and those for the PAF network were 251 and 273 keV. Clearly, the RMS errors on such data configuration do not fully reflect the poor extrapolation performance of DNNs.

%indicating that such a configuration does not give a clear measure for extrapolation performance. 
%We found that the same DNN and PAF networks trained and tested using such data configuration gave no clear measure of extrapolation performance: RMS errors ($N,Z>0$) of inter- and extrapolation for ConvNet are 426 and 446 keV, and those for the PAF network are 251 and 273 keV. 

%AME2020-AME2012 data

\section{Discussion}

There are several potential ways to enhance the current mass model. Novel inputs and activation functions can be explored, along with variations in the number of layers or nodes. Additionally, alternative ideas for applying the penalties during training may enhance performance---not only through different formulations of boundary or Garvey-Kelson constraints, but also by incorporating more physics to guide predictions in the extrapolation regions. 

%Still, there could be limitation as the model is based on the available data. If there is an unknown feature that has no relation to the current data, the model may fail to extrapolate such behavior. 

The effectiveness and utility of physics-related activation functions should be investigated more fully. Tuning the PAF network requires many trials of various sets of activation functions. While a natural approach is to include all possible functions and let the $L_0$ regularization prune weights connected to the unnecessary functions, we found that this approach is typically ineffective: including such functions normally degrades the performance. Additionally, increasing the sparsity or making a much smaller network to find a simple formula (as done in equation learning) also degrades the performance, at least in the case of nuclear mass predictions. Improvements in sparse learning could solve these issues, with an easier and better usage of PAF networks. The idea of PAF can also be applied to different types of network structures, such as convolutional neural networks or Transformers \cite{A.Vaswani2017}.

%We note that the current study has potential for improvements in various aspects. 

%No 184 and reasons why

\section*{Acknowledgements}

This work was supported by the National Research Foundation of Korea (NRF) grants funded by the Korea government (MSIT) (Grants No. RS-2024-00338255). This work was also supported in part by the U.S. Dept. of Energy (DOE), Office of Science, Office of Nuclear Physics under contract DE-AC05-00OR22725. Computational works for this research were performed partly on the data analysis hub, Olaf in the IBS Research Solution Center.

%\bibliographystyle{elsarticle-num}

%\bibliography{ref}{}

%\iffalse

%\fi

\end{document}